# Hidden magnetic excitation in the pseudogap phase of a model cuprate superconductor

Y. Li[1,†], V. Balédent[2], G. Yu[3], N. Barišić[4,5], K. Hradil[6], R.A. Mole[7], Y. Sidis[2], P. Steffens[8], X. Zhao[4,9], P. Bourges[2], M. Greven[3,*]

1. Department of Physics, Stanford University, Stanford, California 94305, USA
2. Laboratoire Léon Brillouin, CEA-CNRS, CEA-Saclay, 91191 Gif sur Yvette, France
3. School of Physics and Astronomy, University of Minnesota, Minneapolis, Minnesota 55455, USA
4. T.H. Geballe Laboratory for Advanced Materials, Stanford University, Stanford, California 94305, USA
5. 1. Physikalisches Institut, Universität Stuttgart, 70550 Stuttgart, Germany
6. Institut für Physikalische Chemie, Universität Göttingen, 37077 Göttingen, Germany
7. Forschungsneutronenquelle Heinz Maier-Leibnitz, 85747 Garching, Germany
8. Institut Laue Langevin, 38042 Grenoble CEDEX 9, France
9. State Key Lab of Inorganic Synthesis and Preparative Chemistry, College of Chemistry, Jilin University, Changchun 130012, China
† Present address: Max Planck Institute for Solid State Research, 70569 Stuttgart, Germany

**The elucidation of the pseudogap phenomenon of the cuprates, a set of anomalous physical properties below the characteristic temperature $T^*$ and above the superconducting transition temperature $T_c$, has been a major challenge in condensed matter physics for the past two decades [1]. Following initial indications of broken time-reversal symmetry in photoemission experiments [2], recent polarized neutron diffraction work demonstrated the universal existence of an unusual magnetic order below $T^*$ [3,4]. These findings have the profound implication that the pseudogap regime constitutes a genuine new phase of matter rather than a mere crossover phenomenon. They are furthermore consistent with a particular type of order involving circulating orbital currents, and with the notion that the phase diagram is controlled by a quantum critical point [5]. Here we report inelastic neutron scattering results for $HgBa_2CuO_{4+\delta}$ (Hg1201) that reveal a fundamental collective magnetic mode associated with the unusual order, and that further support this picture. The mode's intensity rises below the same temperature $T^*$ and its dispersion is weak, as expected for an Ising-like order parameter [6]. Its energy of 52-56 meV and its enormous integrated spectral weight render it a new candidate for the hitherto unexplained ubiquitous electron-boson coupling features observed in spectroscopic studies [7-10].**

Inelastic neutron scattering (INS) is the most direct probe of magnetic excitations in solids. In the present work, we employed both spin-polarized and unpolarized INS measurements. The use of spin-polarized neutrons was crucial to unambiguously identify the magnetic response reported here, because such neutrons are separately collected according to whether their spins have or have not been flipped in the scattering process, which renders magnetic and nuclear scattering clearly distinguishable (Supplementary Information Section 1). Our measurements were carried out on three samples made of co-aligned crystals, which were grown by a self-flux method [11] and free from substantial macroscopic impurity phases and inhomogeneity (SI Section 2). Hg1201 exhibits the highest value of $T_c$ of all cuprates with one copper-oxygen plane per unit cell, has a simple tetragonal structure, and is furthermore thought to be relatively free of disorder effects [12,13]. The scattering wave vector is quoted as $\mathbf{Q} = H\mathbf{a}^* + K\mathbf{b}^* + L\mathbf{c}^* \equiv (H,K,L)$ in reciprocal lattice units (r.l.u.). Neutron intensities are presented in normalized units in most figures to facilitate a direct comparison of the intensity among the measurements (SI Section 3).

Spin-polarized INS data (Fig. 1) demonstrate the existence of a magnetic excitation throughout the two-dimensional (2D) Brillouin zone in a nearly-optimally-doped sample ($T_c = 94.5 \pm 2$ K, denoted as OP95). Energy scans in the spin-flip channel reveal a resolution-limited feature at low temperatures, with a weak dispersion and a maximum of 56 meV at the 2D zone-corner $\mathbf{q}_{AF}$ ($H = K = 0.5$). The feature cannot be due to a polarization leakage from the non-spin-flip channel (SI Section 4), and it disappears in the spin-flip channel at 300 K (Fig. 1a). Background intensity at 10 K has been measured separately using a combination of different spin-polarization geometries (SI Section 1) and agrees with the intensity at 300 K within the error (Fig. 1a, $H = K = 0.2$). These prove the magnetic origin of the peak at 10 K.

The dispersion of the excitation along [$H,H$] for both this optimally doped sample and an underdoped sample ($T_c = 65 \pm 3$ K, UD65, see SI Section 5), measured with both polarized and unpolarized neutrons, is displayed in Fig. 1c. The weak dispersion (<10%) and the strong response at the 2D zone center $q = 0$ drastically differ from the characteristics of the well-known antiferromagnetic response near $\mathbf{q}_{AF}$ [14,15]. Remarkably, the dispersion of the excitation, which is already present well above $T_c$, reaches its maximum at the same point in energy-momentum-space as the so-called magnetic resonance [16], which in OP95 clearly occurs only below $T_c$ [17]. This is further demonstrated in Fig. 2a: compared to the measurement above $T_c$, substantially higher intensity is observed at $\mathbf{q}_{AF}$ below $T_c$.

We emphasize that the magnetic signal far away from $\mathbf{q}_{AF}$ cannot be attributed to a resonance peak that is broad in momentum, for the following reasons. First, at optimal doping, the temperature dependence of the signal away from $\mathbf{q}_{AF}$ (Fig. 3a) is distinctly different from that of the resonance [17] (Fig. 2a inset). Second, the excitation energy near $q = 0$ differs from that at $\mathbf{q}_{AF}$ (Figs. 1 and S4). Third, the profile of momentum scans at the resonance energy is not symmetric about $\mathbf{q}_{AF}$, but better described by a broad peak centered at $q = 0$ plus a narrower peak centered at $\mathbf{q}_{AF}$ (Fig. S8a). Fourth, the resonance peak in momentum scans does not extend below $H = K = 0.3$ (Figs. S8b-d). Therefore, a magnetic excitation branch in addition to the resonance is required to describe the data, as illustrated in Fig. 1c. This excitation branch is also distinctly different from the well-known 'hourglass' excitations [18,19]: the latter only exist in a limited momentum range near $\mathbf{q}_{AF}$ (the hatched area in Fig. 1c) and only become clearly incommensurate below $T_c$ in $YBa_2Cu_3O_{6+\delta}$ (YBCO) [15], whereas the former is observed all the way to $q = 0$ and above $T_c$ (Fig. 1b and Fig. 3). Moreover, following the notion that the hourglass excitations are collective modes below the electron-hole continuum, they are expected only near $\mathbf{q}_{AF}$ and cannot continuously disperse to $q = 0$ [19].

After the magnetic nature of the excitation was verified with polarized neutrons, further quantitative measurements were carried out with unpolarized neutrons to benefit from the much higher neutron flux. Following standard procedure to extract a magnetic signal [19], phonons and spurious contributions were either removed by subtracting background obtained at high temperature (SI Section 6), or avoided by carefully choosing the measurement conditions (SI Section 5). Measurements at 2D **Q**-positions similar to those in Figs. 1a-b, shown in Figs. S4-6, confirm and extend the spin-polarized results. The excitation also was observed at $\mathbf{q} = (0.5,0)$ (Fig. 2b) and $\mathbf{q} = (0,0.5)$ (Fig. 2c), which are equivalent 2D **Q**-positions rotated 45º away from those summarized in Fig. 1c. The energy width of the excitation was found to remain resolution-limited when measured with better energy resolution (Fig. 2d), which indicates that it is a long-lived mode. Since the excitation is observed at all of those **Q**-positions summarized in the inset of Fig. 2d, we conclude that it is present throughout the entire 2D Brillouin zone.

The temperature dependence of the excitation is best measured away from $\mathbf{q}_{AF}$ and with unpolarized neutrons (SI Section 6). The results are summarized in Fig. 3a. The onset temperature of the excitation $T_{ex}$ is shown in Fig. 3b together with $T^*$ determined from in-plane resistivity [12,20] (SI Section 7) and the onset temperature of the $q = 0$ magnetic order measured by polarized-neutron diffraction [4]. The good agreement among these results suggests that the excitation is a fundamental collective mode of the universal $q = 0$ pseudogap order [3,4]. Since the high-$T_c$ cuprates are not ferromagnetic, a 'decoration' of the unit cell with a net cancellation of moments is required to account for the observed behavior. Consequently, the collective mode can not be understood with conventional $t$-$J$ and one-band Hubbard models [21], which reduce the problem to one site per unit cell, and instead an extended multi-band approach appears necessary [5,22,23].

The unusual phase diagram of the cuprates has been argued to be controlled by an underlying quantum critical point that marks the termination of a distinct order parameter [5,24-26]. A leading candidate is the $q = 0$ magnetic order that preserves the translational symmetry of the lattice [3,4] and which would naturally give rise to excitations centered at $q = 0$. Indeed, such an ordered state involving circulating charge currents has been predicted theoretically [5,27]. On the basis that this current-loop order is describable by an Ising-like Ashkin-Teller model, a rather unique magnetic excitation spectrum with nearly dispersion-free excitations is expected from the discrete symmetry of the order parameter [6], consistent with our findings.

In cuprates, anomalies in the charge excitation spectrum are usually discussed in terms of a coupling between electrons and bosonic modes (phonons or antiferromagnetic spin fluctuations). The hitherto unobserved excitation found here at the same energy as the resonance, but up to higher temperature and all the way to $q = 0$, is a new candidate for the mysterious electron-boson-interaction features observed by photoemission [7], optical spectroscopy [8,9] and scanning tunneling spectroscopy [10]. At $q_{AF}$, the strength of the excitation is comparable to that of the resonance (in Hg1201 (Fig. 2a) and YBCO [19]). While the latter is located at $q_{AF}$, the former extends throughout the entire Brillouin zone (Fig. 2). Since the area of resolution ellipsoid in 2D momentum space is a few percent of the Brillouin zone, we estimate that the momentum-integrated spectral weight of the excitation branch is at least an order of magnitude greater than that of the resonance, and comparable to that of the full antiferromagnetic response in underdoped YBCO (the integrated spectral weight between 25 and 100 meV is believed to be several times larger than that of the resonance) [15,28]. In other words, about half of the total magnetic spectral weight is located within a narrow range around the resonance energy, and has been hidden so far, in part due to the excitation's weak momentum dependence. It remains an open question whether the coincidence of energy scales of the excitation and resonance is accidental, or if there exists a profound physical connection.

All evidence suggests that Hg1201 not only is representative of the cuprates, but even a model compound, and therefore experiments on Hg1201 can be expected to reveal the essence of the underlying physics most clearly. Given the universal existence of the pseudogap phase, of the $q = 0$ magnetic order, and of the electron-boson coupling features in the 50-60 meV range, we expect the excitation branch to be present in other hole-doped cuprates as well (SI Section 9).

**Acknowledgements** We thank T. H. Geballe, S. A. Kivelson, E. M. Motoyama, and C. M. Varma for valuable discussions. This work was supported by the Department of Energy and National Science Foundation, and by the National Natural Science Foundation, China. Y.L. acknowledges support from the Alexander von Humboldt Foundation during the final stage of completing the manuscript.

**Author Contributions** M.G., P.B., and Y.L. planned the project. Y.L., V.B., and G.Y. performed the neutron scattering experiments. Y.L., N.B., and X.Z. characterized and prepared the samples. N.B. performed the resistivity measurements. P.S., R.A.M., K.H., Y.S. and P.B. were local contacts for the neutron scattering experiments. Y.L. and M.G. analyzed the data and wrote the manuscript.

**Author Information** Correspondence and requests for materials should be addressed to M.G. (greven@physics.umn.edu).

**Figure 1 | Identification of a weakly-dispersing magnetic collective mode.** (**a**) Spin-flip energy scans for sample OP95. Background (empty squares) is measured at 10 K by a method described in SI Section 1 and approximated together with the data at 300 K by a parabolic baseline (red line). The 10 K data are fit to a Gaussian (blue line) on this baseline, with a small offset to account for the possible background change with temperature. Similar baselines are used in Figs. 1b and 2a. Horizontal bar indicates instrument energy resolution of (FWHM). (**b**) Spin-flip energy scans at additional **Q**-positions, offset for clarity. (**c**) Summary of dispersion along [H,H]. Different symbols represent measurements using different spectrometers: IN20 (circles), PUMA (squares), 2T (triangle) and IN8 (reversed triangle). The measurement on spectrometer IN20 is spin-polarized; all others are unpolarized. Data are presented in Figs. 1a-b, 2c-d and S4-6. Conventional magnetic response near $\mathbf{q}_{AF}$ (vertical dashed line) is present in the hatched area (estimated based on Figs. S8b-d), where the determination of the dispersion using energy scans may be less accurate. Error bars represent statistical and fit uncertainties (one standard deviation).

**Figure 2 | Presence of the collective mode throughout the entire 2D Brillouin zone.** (**a**) Spin-flip energy scans at $\mathbf{q}_{AF}$ below and above $T_c$. Arrow and dashed line indicate the estimated intensity change due to the resonance. The inset is adapted from Ref. 17 and illustrates the intensity change of the resonance below $T_c$. (**b**) Unpolarized measurement for sample UD89 ($T_c = 89 \pm 3$ K) at a **Q**-position away from the 2D zone diagonal. (**c**) Unpolarized measurements for sample UD65 at three different **Q**-positions in the first 2D Brillouin zone. (**d**) Unpolarized measurement for sample UD65 using better energy resolution. A constant has been subtracted from the 403 K data for better comparison. The peak at 5 K is no longer present at 403 K well above $T^*$. Lines in all panels are Gaussian fits which serve as guides to the eye. Empty symbols in (d) indicate measurement points that seem contaminated by phonons (below 50 meV) and a spurious contribution (at 53 meV). Inset of (d) summarizes the **Q**-positions (color-coded for the main panels) at which the measurements were performed. Horizontal bars indicate energy resolutions of the instruments (FWHM) and error bars represent statistical uncertainty (one standard deviation).

**Figure 3 | Temperature dependence of the collective mode demonstrates its connection to the pseudogap phenomenon.** (**a**) Temperature dependence of intensity measured at 53 meV, **Q** = (0.2,0.2,5.2) for sample OP95 (triangles) and at 54 meV, Q = (0.2,0.2,6.0) for sample UD65 (circles), after background subtraction (SI Section 6) and being normalized to values at the lowest temperature. Lines are empirical power-law fits (SI Section 6). The onset temperature $T_{ex}$ is $211 \pm 13$ K near optimal doping and becomes considerably higher ($335 \pm 23$ K) at the lower doping, with no abrupt change near $T_c$ in both cases. (**b**) Summary of characteristic (onset) temperatures. Red circles: excitation branch (this work); blue squares: $q = 0$ magnetic order [4]; green triangles: in-plane resistivity deviation ([12,20] and SI Section 7). Hole concentrations are determined after Ref. 29 based on the doping dependence of $T_c$ in Hg1201 (black line). Error bars represent one standard deviation.

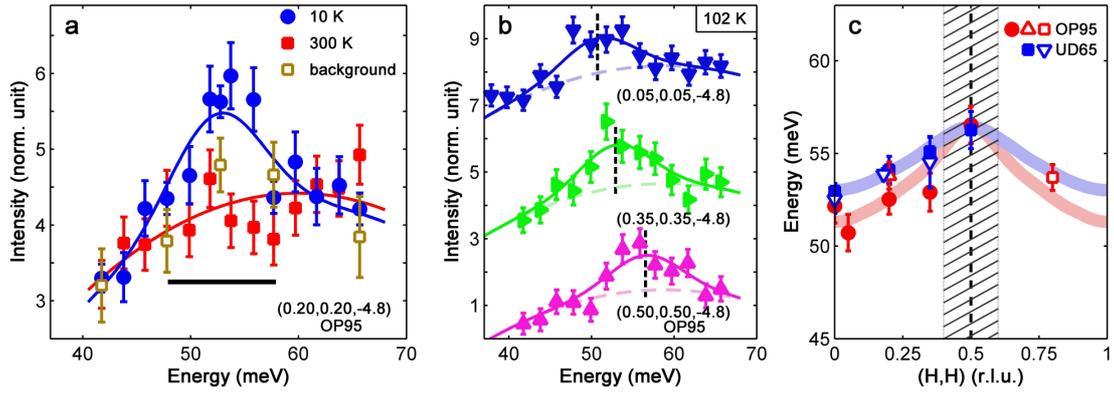

Figure 1

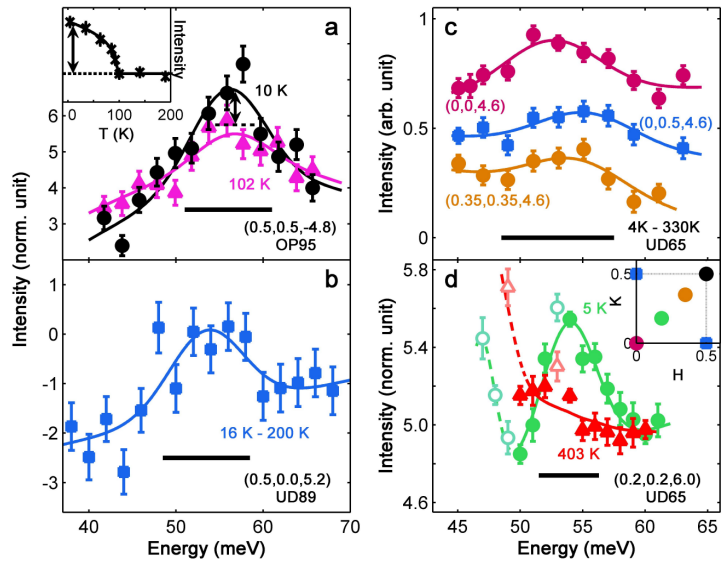

Figure 2

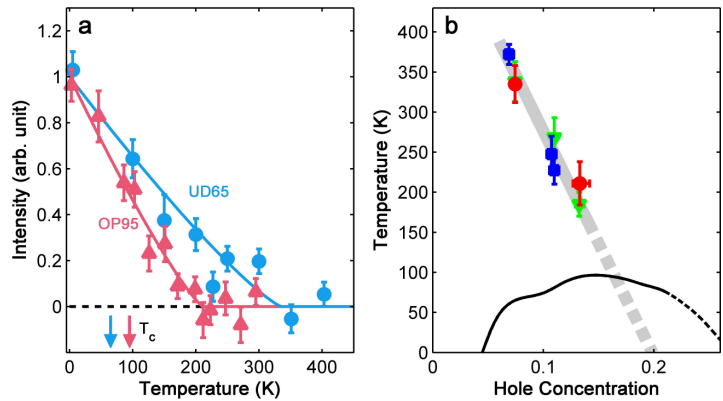

Figure 3

# Supplementary Information

## S1. Neutron scattering methods

In our spin-polarized experiment, the spin polarization of incident neutrons can be freely chosen. The primary spin geometry is with incident polarization (**S**) parallel to momentum transfer (**Q**), in which magnetic and nuclear scattering occurs only in the spin-flip and non-spin-flip channels, respectively. The unwanted nuclear scattering is suppressed by a factor of 15 and becomes practically undetectable. Two additional polarization geometries were used to determine the genuine background level. The method employs the principle that neutrons scattered in the spin-flip channel probe magnetic response perpendicular to both **S** and **Q**. The sum of the intensities ($I$) in the **S**⊥**Q** (polarization perpendicular to **Q** and in the horizontal scattering plane) and **S**∥**Z** (polarization vertical) geometries should contain the same magnetic signal as in the **S**∥**Q** geometry, but twice the background. Consequently, the quantity $I_{S\perp Q} + I_{S\parallel Z} - I_{S\parallel Q}$ represents only the background (see Fig. 1a).

The spin-polarized experiment was carried out on spectrometer IN20 at the Institute Laue Langevin (ILL), using a Heusler-Heusler setup (resolution ~10 meV FWHM in the 50–60 meV energy transfer range). Data are presented in Figs. 1a-b, 2a, S3, and S8a. The unpolarized experiments were performed on 2T (Figs. 2b, 3a, and S6) at the Laboratoire Léon Brillouin (LLB), on PUMA (Figs. S4-5 and S8b-d) at the Forschungs-Neutronenquelle Heinz Maier-Leibnitz (FRM II), and on IN8 (Figs. 2c-d and 3a) at ILL. Pyrolytic graphite (PG) analyzers were used in the unpolarized experiments. Together with PG(002) or Cu(200) monochromators, the energy resolution in the 50–60 meV energy transfer range is ~10 meV FWHM or ~5 meV FWHM, respectively. PG filters were used in all measurements to suppress harmonic contamination in the final neutron beam, the energy of which was fixed at 30.5 (Figs. 2c, S4-5, S8b-d) or 35 meV (in all other figures).

## S2. Sample preparation

The Hg1201 single crystals used in this study were grown by a flux method [11] and heat-treated to the desired doping levels [12]. The total mosaic spread of the two primary samples, OP95 (mass ~ 2.0 g) and UD65 (~ 1.8 g), is ~2° FWHM. A third, slightly underdoped crystal (UD89, ~ 1.0 g, mosaic < 0.5°) was studied once in the [$H$,0,$L$] plane (Fig. 2b), and was subsequently heat-treated and became part of OP95.

The $T_c$ values of the crystals were measured piece-by-piece using a Quantum Design MPMS instrument (Fig. S1). Quoted $T_c$ values and errors are defined by mid-points and widths of superconducting transitions. Based on our characterization results (magnetometry and X-ray Laue diffraction), we can rule out the presence of major impurity phase in our samples. We estimate an upper bound of 30 mg of non-Hg1201 phase in sample OP95 (mainly the white Ba-Cu-O phase on the surface of the two largest crystals in Fig. S1), which is less than 2% of the total mass. No noticeable impurity phase was found in sample UD65. Since both the resonance peak at $\mathbf{q}_{AF}$ and the excitation branch (at and away from $\mathbf{q}_{AF}$) were consistently observed in all samples, and given the small volume fraction of the impurity phase and the very large total magnetic spectral weight, the magnetic signal cannot come from the impurity phase. Impurity scattering would also fail to explain the observed temperature dependence that changes with doping and the physically sensible dispersion.

## S3. Converting intensities into normalized units

The measurements in this study were carried out on a number of different spectrometers. To compare the intensities in different measurements, we employed a calibration procedure based on Bragg peak intensities measured under the same condition as the inelastic experiments. The amplitude of a Bragg reflection (*H*,*K*,*L*) is:

$$I = C \cdot R_0(H,K,L;\omega=0) \cdot F_N^2(H,K,L;\omega=0),$$

where *C* is a constant depending on the sample mass and the neutron flux, $F_N$ is the nuclear structure factor, and $R_0$ is related to the resolution function. $R_0$ and $F_N$ can be calculated according to instrument configuration and crystal structure, respectively.

Based on the results in Fig. S2, we estimate that the ratio among the $C \cdot R_0$ values for the three spectrometers is 1:7:21 (IN20: 2T: IN8). In the inelastic measurements, the quantity $C \cdot R_0 \cdot V$ is proportional to the measured intensity of a feature that is weakly dispersing and sharp in energy, where *V* is the volume of the resolution ellipsoid in momentum space and $R_0$ needs to be re-calculated for the inelastic condition.

Throughout the manuscript and this SI, the vertical scales in normalized units correspond to a monitor number which takes one minute of measurement time at the energy transfer of 56 meV. We use IN20 as the standard for the conversion to normalized units. Based on Fig. S2 and our resolution calculation, the raw intensities measured on IN8 (using a Cu(200) monochromator, Fig. 2d) and 2T need to be divided by 40 and 7, respectively. On PUMA, due to radiation-protection restrictions, the beam width in the elastic condition has to be reduced from that at high energy transfers, which makes the calibration unfeasible. Nevertheless, given its similar instrument design as 2T and a 40% higher power of the reactor with a compact-core design, we estimate a conversion factor of 10-15 for PUMA. Since actual measurements may involve complications that are not captured by the resolution calculation, we estimate an error of up to 50% in the conversion, which is about the largest percentage discrepancy among the four curves in Fig. S2.

## S4. Non-spin-flip measurements for sample OP95

In order to verify that the peaks in Figs. 1a-b are not due to phonon scattering 'leaked' into the spin-flip (SF) channel, scans were carried out in the non-spin-flip (NSF) channel at 10 K (Fig. S3). Unlike in the SF channel, no systematic feature is observed. Given the high flipping ratio (~15), we conclude that phonon 'leakage' cannot be the cause for the peaks in Figs. 1a-b.

## S5. Unpolarized measurements of the dispersion

The measurements were performed on PUMA, with the final neutron energy fixed at 30.5 meV. Energy scans were performed at signal positions $\mathbf{Q}_s = (H,H,6.4)$ for sample UD65. The raw data (filled symbols in Fig. S4a, offset for clarity) clearly show a dispersing feature in the 50-60 meV range, consistent with the results for OP95 (Fig. 1a). To further rule out possible spurious effects and to estimate the background, these scans were complemented with reference scans at background positions $\mathbf{Q}_b = (H',H',0)$ (empty symbols in Fig. S4a). $\mathbf{Q}_b$ was chosen such that $|\mathbf{Q}_s| = |\mathbf{Q}_b|$, in order to minimize the difference in powder scattering and instrumental background between $\mathbf{Q}_s$ and $\mathbf{Q}_b$. Since the $\mathbf{Q}_b$ values are distributed well within the FWHM of the $\mathbf{Q}$-resolution, the reference scans were combined to further reduce statistical uncertainty, after confirming that there is no substantial difference among them (Fig. S4b). The dispersion in Fig. 1c is obtained by fitting the background-subtracted data (Fig. S4c) to resolution-limited Gaussian peaks with a common baseline.

For part of sample OP95, a measurement of the excitation away from $\mathbf{q}_{AF}$ was made at $\mathbf{Q} = (0.8,0.8,4.8)$ (Fig. S5). Intensity difference between low and high temperatures reveals the magnetic signal (see Ref. 17 for the raw data). A polarized measurement at this $\mathbf{Q}$-position was inconclusive due to the weakness of magnetic signal.

## S6. Unpolarized measurements of temperature dependence

The measurement for sample OP95 was performed on 2T. The magnetic excitation away from $\mathbf{q}_{AF}$ is most clearly revealed by the intensity difference between low and high temperatures (Fig. S6b), a method that is commonly employed in neutron scattering studies [19,30-32]. The temperature dependence was measured at the peak position (53 meV) and two candidate background positions (45 and 61 meV) (Fig. S6c). It was then noticed that the intensity at 45 meV increased rapidly at high temperature and was likely affected by phonons. This was confirmed by a careful inspection of the data in Fig. S6a: the blue dashed line (3 K) goes through both candidate background points, whereas the red dashed line (same slope, 295 K) goes through the points at 53 and 61 meV and misses all the points near 45 meV. Therefore, only 61 meV is used as the background. Fitting the intensity difference between 53 and 61 meV to an empirical power-law gives $T_{ex} = 211 \pm 13$ K (Fig. S6d).

The temperature dependence for sample UD65 (Fig. 3a) was similarly measured at $\mathbf{Q} = (0.18,0.18,6.0)$ at 50, 54 and 60 meV (on IN8, full scans at the lowest and highest temperatures are displayed in Fig. 2d). A power-law fit gives $T_{ex} = 335 \pm 23$ K. For both fits in Fig. 3a, the power-law exponents were empirically fixed to 1.2.

## S7. *T\** determined by *ab*-plane resistivity

Resistivity data were reported in Ref. 12 ($T_c$ = 81 K) and Ref. 20 ($T_c$ = 94 K). Here we show additional results for an underdoped sample ($T_c$ ~ 65 K, Fig. S7). We follow the widely-accepted convention of extracting $T^*$, defined as the temperature below which the in-plane resistivity deviates considerably from an approximately linear behavior at higher temperatures. We estimate a conservative uncertainty of 25 K on $T^*$.

## S8. The resonance at $q_{AF}$ in Hg1201

Additional data for the resonance are presented in Fig. S8. The resonance peak is centered at $q_{AF}$ and does not extend to $q = 0$ (momentum width consistent with those in YBCO and Tl2201, see discussion in Ref. 17). At energies above and below 56 meV, a weak signal centered at $q_{AF}$ is observed, which is similar to the situation near the resonance in other cuprates, and which might be the counterpart of the 'hourglass' excitations in Hg1201. The limited statistical accuracy does not allow us to determine whether the response below and above the resonance energy is commensurate.

## S9. How could the excitation branch have been missed so far?

To the authors' knowledge, there exist no reports in the literature of the cuprate superconductors of spin-polarized INS measurements at momentum positions near the 2D zone center, and especially not of measurements involving energy scans near the 2D zone center. As demonstrated by our results for Hg1201, such scans reveal the excitation branch most clearly. The lack of such measurements is partly due to the fact that the main focus of the previous polarized measurements had been to confirm unpolarized results for the antiferromagnetic response, especially the resonance at $q_{AF}$ [16]. This response is best measured with momentum scans at constant energy [33-35], or with energy scans at $q_{AF}$ [35,36], but not with energy scans near $q = 0$.

Unpolarized measurements, which are much more common, inevitably contain both phonon and magnetic contributions. Phonon signals can be strong, and they may stem not only from the sample, but also from the sample mount (usually made of aluminum). Therefore, it is common practice to ignore 'bumps' in raw data of unpolarized energy scans that have no obvious connection to the response at $q_{AF}$. For the same reason, momentum scans were regarded as more useful, because (weakly-dispersive) optical phonons are expected to only contribute to the 'background.' While this is true when the signal of interest is peaked at specific **Q**-positions (such as $q_{AF}$), the magnetic excitation branch reported here would unfortunately also only contribute to the presumed 'background'. Such 'background' in momentum scans (or in constant-energy slices in time-of-flight measurements) was either removed or received little attention when data at different energies were

compared (for example, see [14,15,19,28,32,37-41]).

One way of extracting magnetic signal from unpolarized data is to compare scans at different temperatures. With increasing temperature, phonon intensities tend to increase due to the Bose factor, whereas magnetic signals tend to decrease. Even though the effects of phonons and of magnetic excitations might cancel each other, there is still a chance to extract the latter if the correct type of scan is used. This has been one of our main approaches to measure the excitation branch using unpolarized neutrons. However, just as for energy scans near the 2D zone center with polarized neutrons, we did not find any comparison in the literature between energy scans at small $H$ and $K$ values at different temperatures. Moreover, the excitation branch reported here sets in near $T^*$, but comparisons that received most attention previously were for temperatures across $T_c$ [19,30,32,34,38,42-44]. Such comparisons were optimized to highlight the resonance, but might have missed a substantial part of the intensity change due to the excitation branch. In fact, this focus on the intensity change across $T_c$ had prevented the community from realizing the importance of the normal-state response near $q_{AF}$ [15,37,45] until long after the resonance was discovered. The situation is similar in our case. A hint of the excitation branch was serendipitously observed when we measured the resonance in Hg1201 [17] and compared data at 10 K and room temperature (rather than just above $T_c$), and this observation motivated the present work. The high resonance energy in Hg1201 has provided an advantage in this regard, because the Bose factor change between 10 K and 300 K for 55 meV is only half of that for 40 meV, the approximate energy of the (odd-party) resonance energy in YBCO. Had the excitation branch been at ~ 40 meV, *i.e.*, close to the odd-parity resonance energy in double-layer systems, it would have been more difficult to discern, because the effect could have been more easily hidden by the increase of phonon scattering due to the Bose factor.

One might think that time-of-flight measurements should have accidentally revealed the excitation branch, because a large portion of the momentum-energy space is mapped out in such measurements without selective focus. However, such measurements have several drawbacks as well. First, the commonly used sample orientation (*c*-axis along beam axis) does not allow access to $H = K = 0$, where the excitation branch gives the strongest signal. Second, the 'background' is hardly understood and generally ignored. Third, the signal intensity at a given momentum-energy position is low, which makes careful studies on the temperature dependence very time consuming. Finally, to date, detailed spin-polarization analysis has not been possible with time-of-flight spectrometers.

The excitation might be particularly difficult to observe in $La_{2-x}Sr_xCuO_4$, one of the most intensively studied cuprates. In this compound, the unusual magnetic order was recently found to be very short-ranged [46], either because of competing stripe order [47] or strong disorder effects [13], which can be expected to lead to highly damped (broad in energy) excitations that are difficult to detect.

**Supplementary figures:**

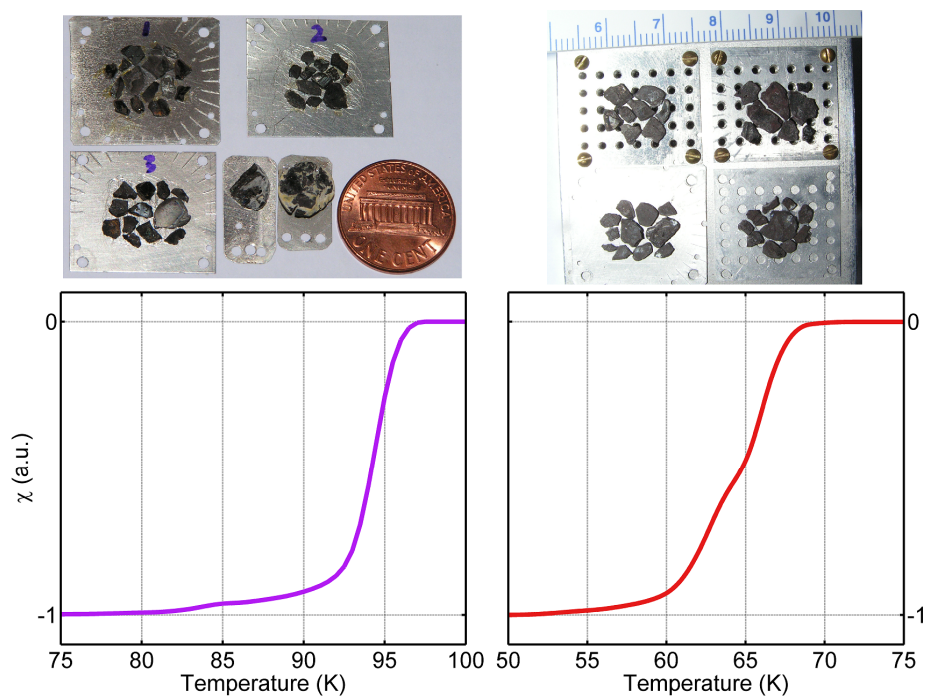

**Figure S1.** Pictures and summed magnetometry curves for the two primary samples: OP95 (left) and UD65 (right).

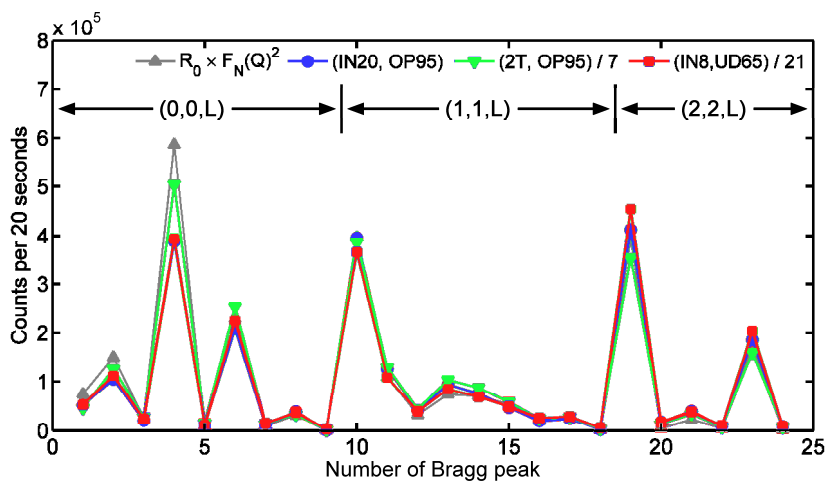

**Figure S2.** Nuclear Bragg peak intensity measured on (001) to (009), (110) to (118) and (220) to (225), labeled as numbers 1 to 24, for three spectrometers. Also plotted are the (rescaled) product of the calculated $R_0$ values for spectrometer IN20 and $F(Q)^2$ for the Bragg reflections. The overall agreement demonstrates that the three spectrometers have similar (and regular) resolution functions, and that the intensity difference can be approximated by the single quantity $C \cdot R_0$. The calculated $R_0$ values of other spectrometers have similar ratios among the different reflections, and are therefore not shown.

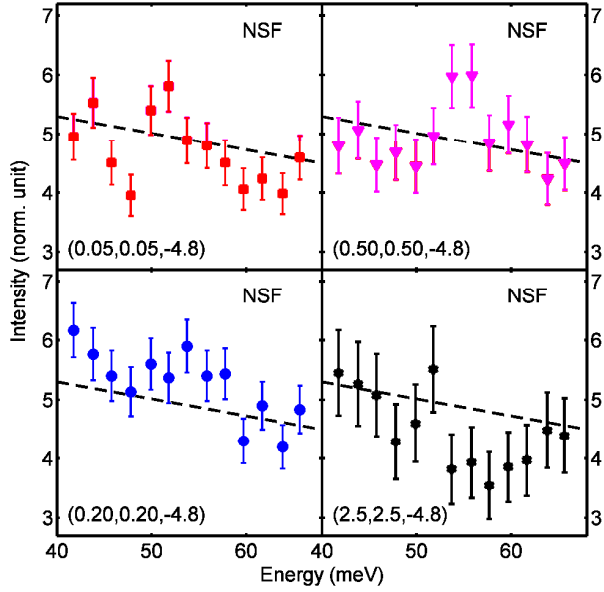

**Figure S3.** Non-spin-flip energy scans for sample OP95 at the same **Q**-positions as in Figs. 1a-b at $T = 10$ K (measured on IN20). An identical dashed line is drawn in all panels to facilitate comparison among the panels. The absence of peak at $H = 2.5$ rules out a phonon interpretation of a possible feature at $H = 0.5$. Error bars indicate statistical uncertainty (one standard deviation).

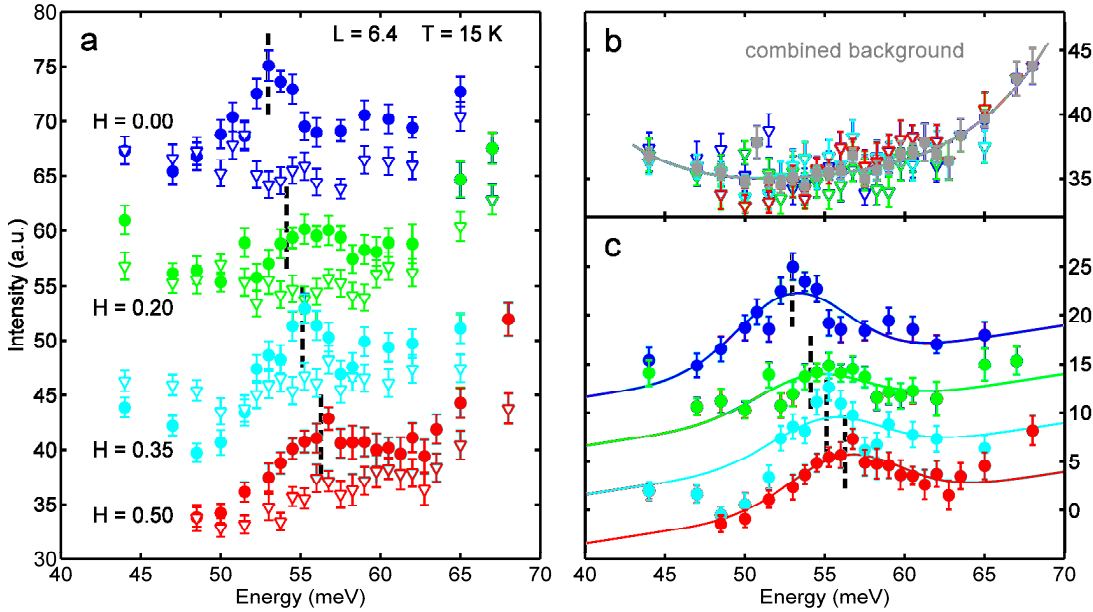

**Figure S4.** (**a**) Unpolarized energy scans for sample UD65 at $(H,H,6.4)$ (filled circles) and corresponding $\mathbf{Q}_b$-positions (empty triangles, see text). The choice of positive $L$ is in accordance with the configuration of the PUMA instrument, so that the resolution ellipsoid has a momentum-energy space orientation similar to that on IN20 with negative $L$. (**b**) The reference scans are combined and described by a polynomial (grey line). (**c**) Background-subtracted data fit by resolution-limited (9.5 meV FWHM) Gaussian. Vertical scales correspond to one minute of measurement time at 56 meV (before conversion to normalized units). Data in (a) and (c) are offset for clarity. Vertical dashed lines indicate the fit peak positions. Error bars indicate statistical uncertainty (one standard deviation).

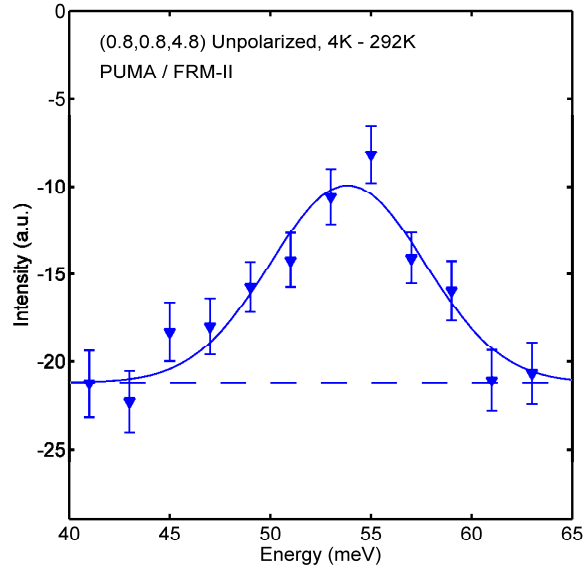

**Figure S5.** Unpolarized measurement on PUMA for part of sample OP95. The solid line is a resolution-limited (9.5 meV FWHM) Gaussian fit to the data. Error bars indicate statistical uncertainty (one standard deviation).

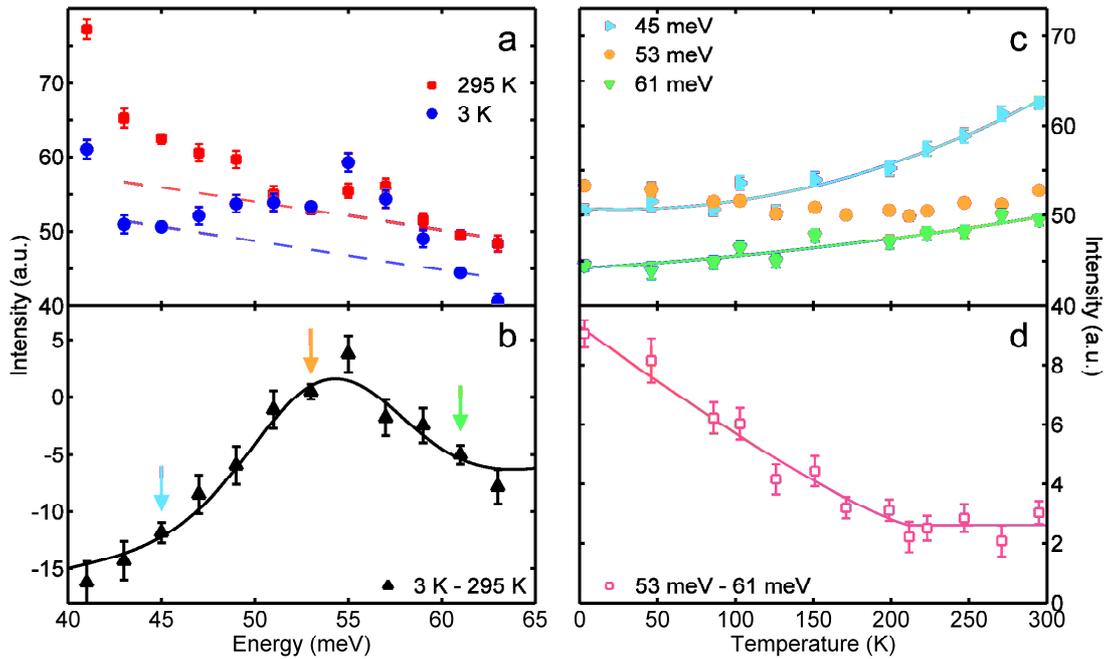

**Figure S6.** (**a**) Unpolarized energy scans at **Q** = (0.2,0.2,5.2) for OP95. The choice of positive $L$ is in accordance with the configuration of the 2T instrument, so that the resolution ellipsoid has a momentum-energy space orientation similar to that on IN20 with negative $L$. The dashed lines are for comparing the background levels and have the same slope (see text). (**b**) Intensity difference between 3 K and 295 K, fit to a resolution-limited (9 meV) Gaussian (center 53.8 ± 0.9 meV) on a sloping baseline. (**c**) Temperature dependencies at the arrowed positions in (b). The lines are quadratic smoothing curves. (**d**) Difference between 53 meV data and the smoothing curve for 61 meV in (c). An empirical power-law fit (solid line) gives an onset temperature of $T_{ex}$ = 211 ± 13 K. Vertical scales correspond to one minute of measurement time at 56 meV (before conversion to normalized units). Error bars indicate statistical uncertainty (one standard deviation).

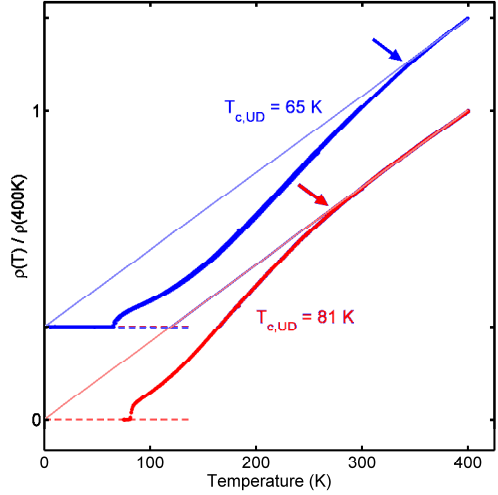

**Figure S7.** In-plane DC resistivity measurements of underdoped Hg1201 single crystals using the standard 4-probe method [12]. The thin lines are linear fits to the data above 300 K and 350 K for the $T_c \sim 81$ K and $\sim 65$ K samples, respectively. Values of $T^*$ (268 K and 338 K) defined by considerable deviation from an approximately linear behavior are indicated by the arrows.

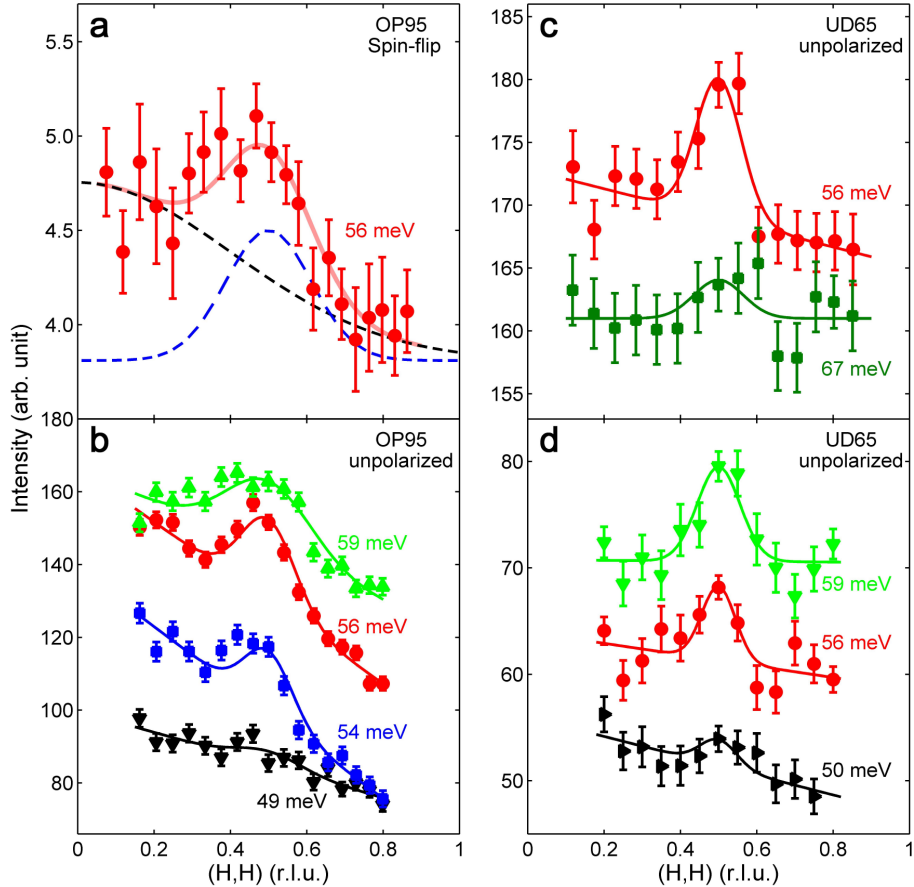

**Figure S8.** Momentum scans at $T < 15$ K. (**a**) Spin-flip data for sample OP95 showing the resonance at $\mathbf{q}_{AF}$ (blue dashed line) together with the excitation centered at $q = 0$ (black dashed line). The scan was performed by rocking the sample (keeping $Q$ constant) through (0.5,0.5,-4.0). (**b**) Unpolarized rocking scans for OP95 at several energies near the resonance. Data are offset for clarity. Adapted from [17]. (**c-d**) Unpolarized momentum scans for UD65. The measurements in the two panels were performed using different configurations. Data are offset for clarity.